\documentclass[final,twocolumn,5p,12pt]{elsarticle}
\usepackage[T1]{fontenc}
\usepackage{graphicx}	
\usepackage{amsmath,amssymb}
\usepackage{times}
\usepackage{units}
\usepackage[english]{babel}
\usepackage[utf8]{inputenc}
\usepackage{booktabs,array}
\usepackage{listings}
\usepackage{xspace}
\usepackage{fp}
\usepackage{ifthen}
\usepackage{color}

\newcommand{\refeq}[1]{Eq. (\ref{#1})}
\newcommand{\reffig}[1]{Fig. \ref{#1}}

\def \br {\mathbf{r}}
\def \S {\mathcal{S}}
\def \T {\mathcal{T}}
\def \R {\mathcal{R}}

\def \n {\mathbf{n}}

\def \be {\begin{equation}}
\def \ee {\end{equation}}
\def \bear {\begin{eqnarray}}
\def \eear {\end{eqnarray}}

\def \D {\hat{D}}
\def \H {\hat{H}}

\def \kap {\boldsymbol \kappa}

\journal{Computational Material Science}
\begin{document}
\begin{frontmatter}

\title{Electronic Structure Trends of M\"obius Graphene Nanoribbons from Minimal-Cell Simulations}

\author{Topi Korhonen}
\author{Pekka Koskinen\corref{cor}}
\ead{pekka.koskinen@iki.fi}
\address{NanoScience Center, Department of Physics, University of Jyv\"askyl\"a, 40014 Jyv\"askyl\"a, Finland}

\cortext[cor]{Corresponding author}

\begin{abstract}
Investigating topological effects in materials requires often the modeling of material systems as a whole. Such modeling restricts system sizes, and makes it hard to extract systematic trends. Here, we investigate the effect of M\"obius topology in the electronic structures of armchair graphene nanoribbons. Using density-functional tight-binding method and minimum-cell simulations through revised periodic boundary conditions, we extract electronic trends merely by changing cells' symmetry operations and respective quantum number samplings. It turns out that for a minimum cell calculation, once geometric and magnetic contributions are ignored, the effect of the global topology is unexpectedly short-ranged. 

\end{abstract}


\end{frontmatter}
%\pacs{71.15.Dx,71.15.-m,73.22.Pr,61.48.Gh}
% 71.15.Dx 	Computational methodology (Brillouin zone sampling, iterative diagonalization, pseudopotential construction)
% 71.15.-m 	Methods of electronic structure calculations
% 73.22.Pr    Electronic structure of graphene
% 61.48.Gh    Structure of graphene

%\maketitle

\section{Introduction}

%PBC: GENERAL RELEVANCE
Bulk crystals are conventionally simulated using translational symmetry and periodic boundary conditions.\cite{bloch} Periodicity means that the electron wave function is symmetric with respect to a certain number of translations along a given dimension of the simulation cell. Since the periodicity occurs for all three dimensions of the cell simultaneously---with all the opposing faces of a parallelpiped intertwined to come into contact with one another---, it cannot represent physical reality; the periodicity is often just a mathematical trick. For some structures, however, the periodicity does represent physical reality, even without the need to consider the limit of infinite system size. Such structures include a ring-like structure due to one-dimensional periodicity, a toroidal structure due to two-dimensional periodicity, and a M\"obius ribbon due to one-dimensional roto-translational periodicity. 
 
M\"obius ribbon forms upon connecting the ends of a half-twist rectangular strip, so it has only one surface and one edge. Its periodicity is fascinating: one must traverse a distance twice the length of the ribbon to return to the starting point. This periodicity has profound implications on electron wave functions, and it has triggered interest for both experimental and theoretical investigations. Experiments have realized M\"obius hydrocarbons\ \cite{ajami,ajami_chemistry_2006} and single crystal M\"obius strips\ \cite{tanda}. Theoretical works have addressed electronic properties of M\"obius graphene\ \cite{wang,guo,jiang,caetano_langmuir_09} or other \cite{ballon,mila,zhao,yakubo} ribbons, as well as classical properties of elastic M\"obius ribbons\ \cite{starostin}. Unfortunately, first-principles calculations are hard to analyze, making trends and their origins difficult to extract.

%Revising Bloch's theorem it may be extented to deal with larger group of symmetries. This leads to the revised periodic boundary conditions \cite{rpbc_pekka, rpbc_oleg}. In the revised periodic boundary conditions we require that the symmetry operations of the system form an Abelian group, that is the symmetry operations commute among themselves. With the revised periodic boundary conditions it is possible to model systems with symmetries beyond translation, for example systems with wedge symmetry. Moreover, the periodic boundary conditions are now actual physical condition for systems with genuine periodicity, for example in the  benzene molecule.       
 
Here, we compare M\"obius graphene nanoribbons to straight ones by using density-functional tight-binding employing revised periodic boundary conditions. The revised periodic boundary conditions enable the extraction of  length- and width-dependent trends that arise from topological effects alone. The approach requires only minimal simulation cells; both topology and length are controlled by parameters external to the cell. We find that the topology has an unexpectedly short-ranged effect in the electronic properties of M\"obius graphene nanoribbons with realistic aspect ratios. While the changes in topology for real ribbons makes big differences for total energies, geometries, and electronic structures, we find that, when excluding the geometrical effects, the M\"obius topology itself has an unexpectedly small effect on ribbons' properties, especially for realistic aspect ratios.

\section{Minimal-cell simulations}

To establish the approach of treating the M\"obius topology, we briefly present the framework of revised periodic boundary conditions.\cite{rpbc_pekka, rpbc_oleg}. Let the potential $V(\br)$ in a non-interacting Hamiltonian $\H = -\nabla^2/2 + \hat{V}(r)$ be invariant under the symmetry operation $\S^{\n} \equiv \S_1^{n_1}\S_2^{n_2}\cdots$, that is,
\be
\label{eq:sv}
\D(\S^{\n})V(\br) \equiv  V(\S^{-\n }\br) = V(\br)
\ee
for a set of commuting symmetry operations $\S_i^{n_i}$. Energy eigenstates $\psi_{a\kap}(\br)$ then satisfy 
\be
\label{eq:DPsi}
\D(\S^{\n})\psi_{a \kap}(\br) = e^{-i \kap \cdot \n}\psi_{a \kap }(\br),
\ee 
with the quantum numbers $\kap=(\kappa_1,\kappa_2,\ldots)$. This framework differs from Bloch's theorem only in choosing generalized symmetry operations $\S_i^{n_i}$ over simple translations. The key ingredient is that, by a clever choice of $\S_i^{n_i}$'s, a minimal simulation cell will suffice to investigate extended structures with custom-made symmetries, periodicities, and even topologies.\cite{koskinen_PRB_10b, kit_PRB_12,koskinen_PRB_12}

Bearing this framework in mind, consider two symmetry operations, $\R$ for a $\pi$-rotation (half-twist) around the $z$-axis and $\T$ for a translation of $l$ in the $z$-direction. For a M\"obius ribbon of length $L=Ml$, the wave function should then satisfy
\be
\label{eq:equiv_sym_op}
\D(\T^M)\psi(\br) = \D(\R)\psi(\br),
\ee
which means that translating a length $L$ along the ribbon equals to a half-twist. We term ribbons modeled this way \emph{topological M\"obius ribbons}, as they include the correct topology, and yet exclude geometric contributions from bending or streching.\cite{starostin,koskinen_APL_11}

\begin{figure}[t]
\begin{center}
	\includegraphics[width= 0.8\linewidth]{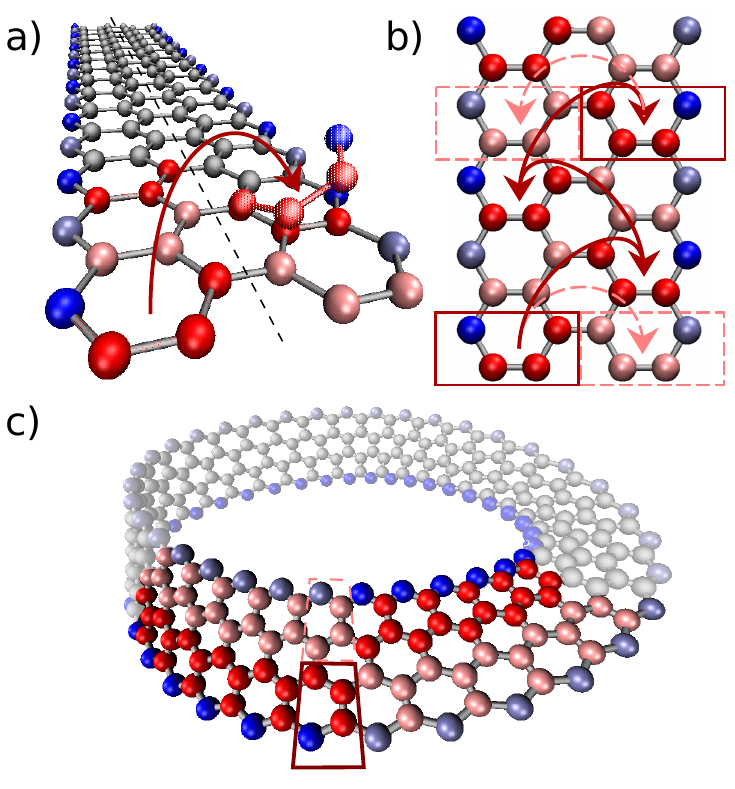} %
\end{center}
	\caption{Periodic boundary conditions of M\"obius ribbons. a) Topological M\"obius ribbon is effectively a straight ribbon, periodicity being imposed only upon the topology of the wave function. b) The $\pi$-rotation $\R$ (dashed arrow) and roto-translation $\R\T$ (solid arrow) performed on a minimal unit cell of a $4$-zigzag graphene nanoribbon. By combining $\T\R$ and $\R$, we may construct the whole ribbon from the minimal cell. Here the M\"obius topology comes through associating the two cells (solid rectangles) that are connected by three operations of $\R\T$ (periodic boundary condition $\T^3 = \R$). c) Illustration of the equivalence of $\T^M = \R$ for a finite, real-space M\"obius ribbon.}
\label{fig:schematic}
\end{figure}

To apply the revised periodic boundary conditions to simulate M\"obius ribbons in practice, we choose two symmetry operations, rotation $\R=S_1$ and roto-translation $\T\R=S_2$. This choice suggests a minimal simulation cell (Fig.\ref{fig:schematic}). For ribbons with symmetric edges, $\S_1^2=\hat{1}$ implies either $\kappa_1=0$ (symmetric state) or $\kappa_1=\pi$ (antisymmetric state). The periodicity along the ribbon is defined by the boundary condition 
\be
\D(\S_2^{M})\psi_{a \kap }(\br)=\D(\S_1^{N})\psi_{a \kap }(\br),
\ee
where $N=0$ or $1$ and the ribbon length $L=M l$. Rearranging the above boundary condition into 
\be
\D(\T^{M})\psi_{a \kap }(\br)=\D(\R^{N-M})\psi_{a \kap }(\br),
\ee
and comparing it with \refeq{eq:equiv_sym_op} reveals that even $N-M$ represents a straight ribbon and odd $N-M$ a M\"obius ribbon. Juxtaposing this boundary condition with \refeq{eq:DPsi} we get $e^{-iM \kappa_2}=e^{-iN \kappa_1}$, which gives the quantum-number sampling for $\kappa_2$ as 
\be
\label{eq:ksampling}
\kappa_2=\frac{N\kappa_1+2\pi m}{M},\;\; m=0,\ldots, M-1,
\ee
where $\kappa_1=0$ or $\pi$, as mentioned. Thus, both the topology (via $N-M$) and the length (via $M$) can be controlled by controlling the boundary conditions, while keeping the simulation cell the same.

The choice of the symmetry operations could be different as well. M\"obius ribbons could be simulated by the symmetry operation $\S=\T\R$ alone, with simulation cell spanning the entire width of the ribbon. This choice, with the boundary condition $(\R\T)^M=\hat{1}$, leads to straight ribbons for even $M$ and M\"obius ribbons for odd $M$. The quantum number sampling is $\kappa=2\pi m/M$ ($m=0,\ldots,M-1$). Unfortunately, this choice cannot be used to simulate M\"obius ribbons for lengths $L=Ml$ when $M$ is even. 

The idea of a "M\"obius boundary condition" has been around for some time. Apart from real, finite M\"obius ribbon simulations, M\"obius or "twisted" periodic conditions have been used for before, especially using a nearest-neighbour tight-binding picture.\cite{mila,wang,yakubo,wakabayashi_JPSC_03} Unlike previous works, with the exception of the recent paper by C\"u\c{c}l\"u \emph{et al.}\cite{Guclu2013}, our minimal-cell setting enables investigating the effect of boundary conditions as a function of length, to extract trends and separate the effect of topology from the geometrical effects. 

Calculations were done by density-functional tight-binding code hotbit, which is equipped with a flexible implementation of custom-made symmetries and boundary conditions. \cite{hotbit_wiki,koskinen_CMS_09} The sole implementation needed was the function for the symmetry operation $\br'=\S_i \br$, along with the corresponding boundary conditions. The quantum-numbers ($\kappa$-point) were sampled according to \refeq{eq:ksampling}. Structures were optimized using the criterion of $10^{-6}$~eV/\AA\ for the atoms' maximum forces.

% Lisatty, tuotu conclusion osiosta ja kirjoitettu uusiksi, pyritaan selvittamaan mika se rakenne ja topologia nyt oikein on
As explained above, we impose the topology only through symmetry operations and boundary conditions. Thus, the ribbon is structurally straight graphene nanoribbon regardless of the topology. In real and finite M\"obius ribbons the strains and bendings would constitute considerable energy contributions already with a single half-twist.\cite{caetano_JCP_08} However, as the purpose of this work is to investigate the effects of the topology alone, the exclusion of the geometrical effects is a sheer advantage.

Finally, we verified the minimum-cell method by comparing the results to corresponding finite M\"obius ribbons. We constructed finite M\"obius ribbons by removing the geometrical variations from the matrix elements. All the electronic properties of the finite M\"obius ribbons and the topological M\"obius ribbons were numerically identical. The minimum-cell method was proven valid.

\section{Electronic structure trends}

\begin{figure}[b!]
\begin{center}
\includegraphics[width= 0.9\linewidth]{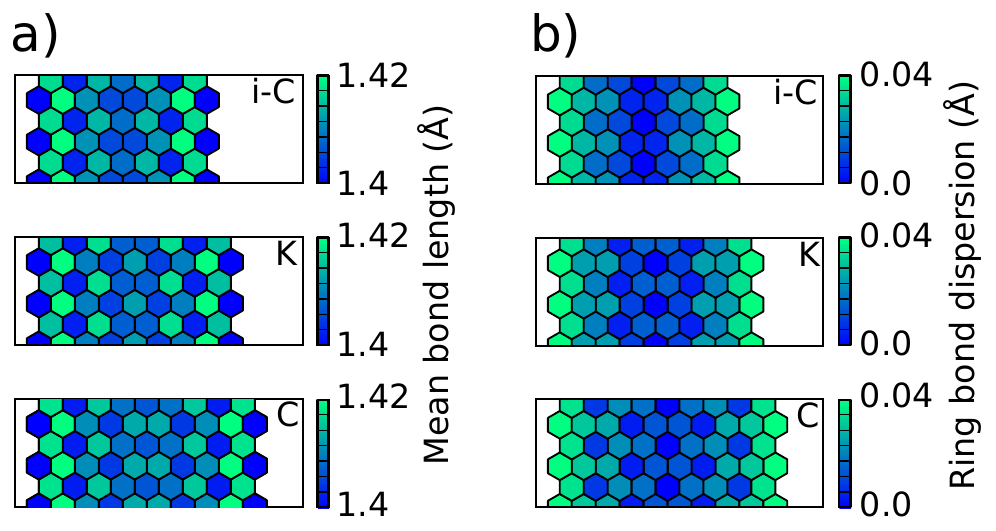}
\end{center}
\caption{\label{fig:structures} Geometric signatures of straight and M\"obius graphene ribbons with three different geometrical structural families, incomplete-Clar (i-C; $W=20.9$ \r{A}), Kekulé (K; $W=23.4$ \r{A}) and Clar (C; $W=25.8$ \r{A}). a) Mean bond lengths of the six bonds in hexagonal rings. b) Dispersions of the bond lengths of the six bonds in hexagonal rings.}
\end{figure}

We performed calculations for hydrogen-passivated armchair graphene nanoribbons with widths ($W$) up to $4.0$~nm, with lengths ($L$) up to $14$~nm, and with both topologies (straight and M\"obius) for all ribbons. Furthermore, we restricted ourselves to ribbons with symmetric edges (symmetric under $\pi$-rotation) in order to single out the effect of topology for a ribbon of given length. Asymmetric edges would have blurred comparisons between straight and M\"obius ribbons due to different number of atoms. This restriction is not a constraint due to the method, but a constraint due to the M\"obius topology. Aspect ratios, however, had no restrictions. While the smallest realistic aspect ratios for M\"obius ribbons lie around $L/W\gtrsim 4$\ \cite{starostin, wang}, the method enabled investigating the effects of topology for even smaller aspect ratios, for the sheer purpose of trend extraction.

%STRUCTURE: Tama jatetaan Hammenyksen valttamiseksi poies

%We begin by presenting the effect of topology on geometric structure. Structure is gauged through two quantities, the mean bond length (the average length of bonds within a hexagon) and the ring bond dispersion (the mean-square deviation of the bond lengths within a hexagon). It turned out that for realistic aspect ratios the topology impacted the relaxed geometry very little (mean bond length deviations $\sim 10^{-6}$~\AA). Such small impact is, of course, relevant only for topological M\"obius ribbons; finite-size effects are discussed later. In what follows, because of this geometrical similarity, we compare properties by using precisely the same relaxed unit cell and atomic positions as they result from calculations for  straight ribbons.

\begin{figure}[t!]
\begin{center}
	\includegraphics[width= 0.8\linewidth]{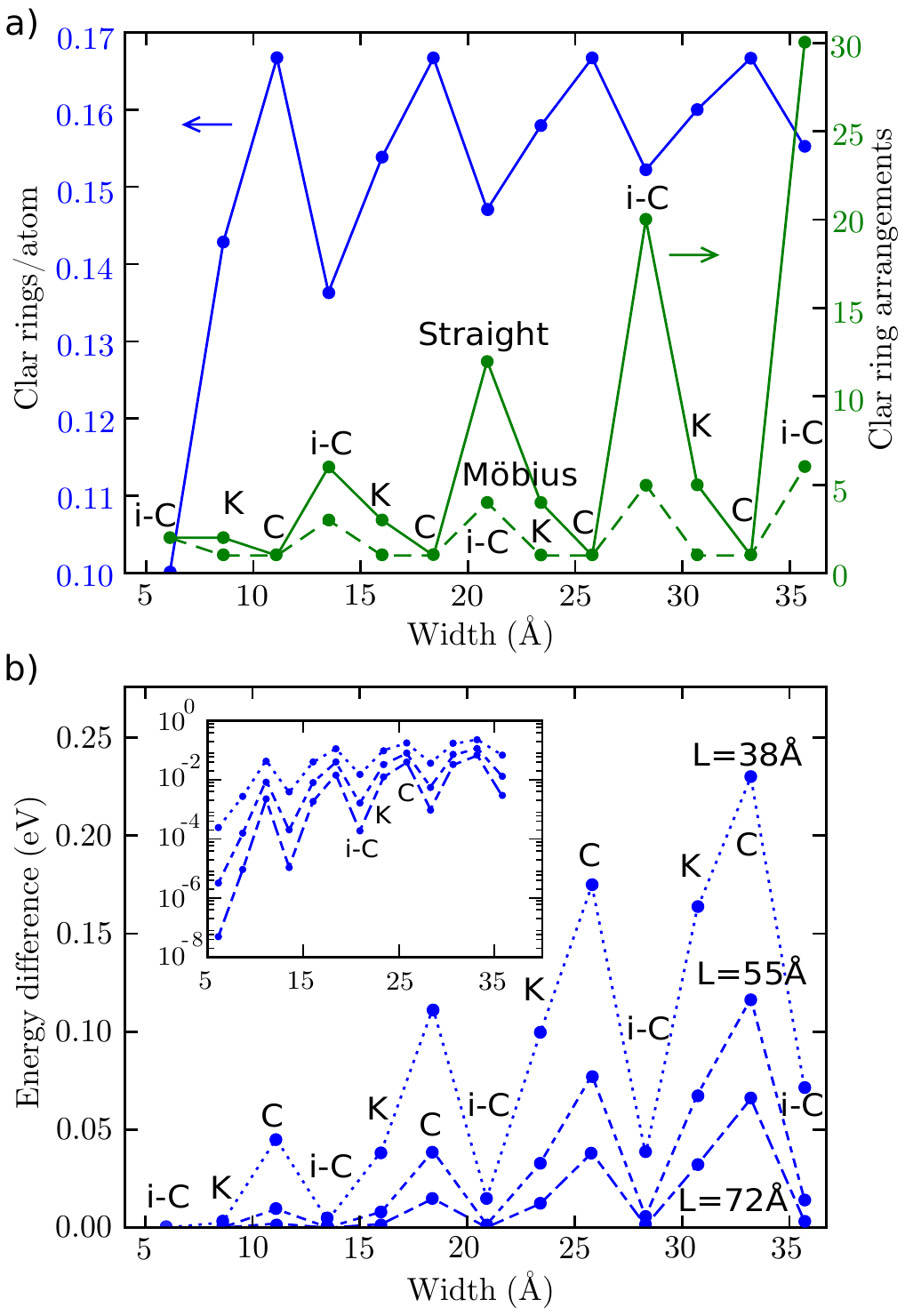}
\end{center}
	\caption{\label{fig:arom_deg} Effect of M\"obius topology on energies. a) The number of Clar rings per carbon atom in the ribbon (same curve for both topologies; left scale). The degeneracy of optimal Clar ring arrangements as a function of ribbon width for straight (solid line) and M\"obius (dashed line) topologies. b) Energy difference $ E_{\text{straight}} - E_{\text{M\"obius}}$ for the two topologies as a function of ribbon width. Symbols indicate structural families. }
\end{figure}

The first observation was that both straight and M\"obius armchair graphene nanoribbons share the three known structural families. Classified by $q=\mod(\mathcal{N},3)$, where $\mathcal{N}$ is the number of atomic rows in an armchair ribbon, they are the so-called Clar (C, $q=0$), Kekul\'e (K, $q=1$) and incomplete-Clar (i-C, $q=2$) structural families (\reffig{fig:structures}).\cite{Martin} The families' geometrical structures, the occurrence of smaller hexagons in certain geometries, and the concomitant changes in ring bond dispersions can be justified by the spatial arrangement of aromatic Clar rings (sextets) in the system.\cite{Martin,wassmann_JACS_10} According to this justification, the Clar's theory for aromatic sextets, the most stable structure and the representation for its electronic structure is found by an arrangement of isolated Clar rings that maximizes their number.\cite{clar} If there are different arrangements with the same number of Clar rings, giving a degeneracy, the resulting electronic structure is presented by a simultaneous combination of these arrangements.\cite{clar} The preservation of structural families implies that the changed topology has a small effect on the patterns of the optimum ring arrangements.

%The degeneracy of optimal arragements depends on the width of the ribbon.  

%We show connection between the total energy difference, the different geometrical structures, presented in \cite{Martin}, and the degeneracy of optimal aromatic ring orderings.  

%However, for small $L/W$ the topology has major effect on the band gap of the system and some impact to the total energy and structure. 

The role of Clar ring arrangements can be illustrated by a look at the energetics (\reffig{fig:arom_deg}b). For a finite ribbon of given length, width, and topology, the arrangement of Clar rings can be calculated explicitly (two adjacent hexagons are never Clar rings simultaneously; therefore in graphene the number of Clar rings per carbon atom attains its maximum value of $1/6$, \reffig{fig:arom_deg}a). Such explicit calculations show that the number of Clar rings in optimal arrangements do not depend on topology, while the arrangement degeneracies do (\reffig{fig:arom_deg}a). When the degeneracies are larger, also the energy differences between different topologies are smaller. This relationship is intuitive: when the arrangement of Clar rings is "rigid", structure will be sensitive to the change in topology (compare Figs. \ref{fig:arom_deg}a and \ref{fig:arom_deg}b). If, however, the arrangement is not rigid but flexible due to a large degeneracy, it will be insensitive to changes in topology. This is why for the incomplete-Clar family, which has the largest degeneracy for the arrangements, the energy changes the least. Clar structures, on the other hand, have only one arrangement of Clar rings, no degeneracy, and are the most sensitive for changes in topology (\reffig{fig:arom_deg}b).

\begin{figure}[t!]
\begin{center}
	\includegraphics[width= 0.8\linewidth]{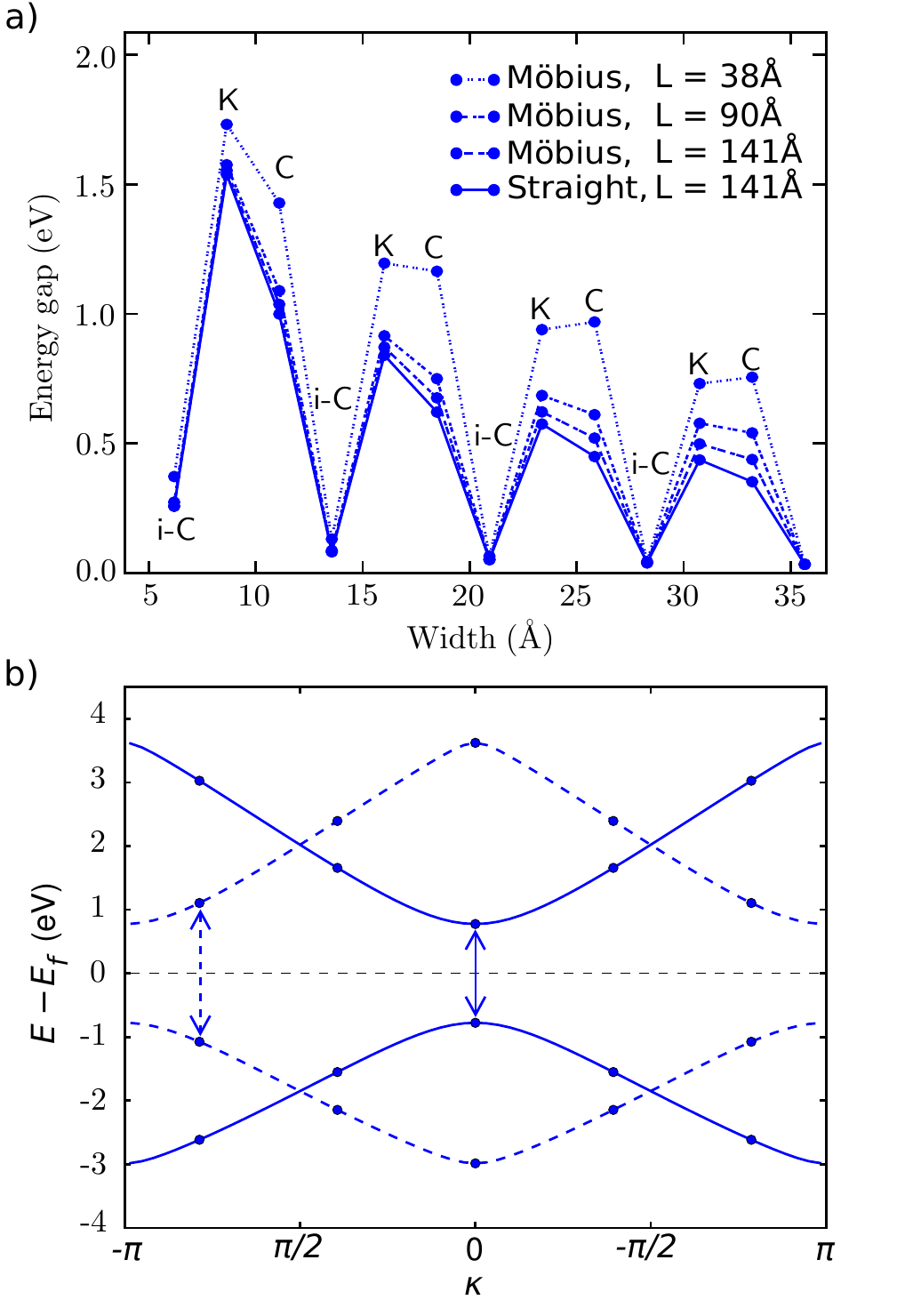}
\end{center}	
	\caption{\label{fig:band_gap} The effect of M\"obius topology on energy gaps. a) Band gaps as a function of ribbon width for different lengths and topologies. b) Band structure of straight and M\"obius ribbons near Fermi-level. If the wavefunctions are odd, the bands of the straight ribbon (solid line) shift by $\pi$ upon changing the topology to M\"obius. Since the $\kappa$-point sampling remains the same (sampled energies with solid spheres), the resulting energy gap changes.}
\end{figure} 

%BAND GAP
Similar arguments are valid also for energy gaps. The gaps of straight ribbons oscillate as a function of width according to the three structural families (\reffig{fig:band_gap}a).\cite{Martin,son_PRL_06} Topology affects the gaps in Clar and Kekul\'e families, but not in the incomplete-Clar family. In the incomplete-Clar family the smallness of the gap is a reminiscent of the large degeneracy of Clar ring arrangements, representing a ``metallic'' electronic structure with its delocalized electron wave functions. Large degeneracy in Clar rings hence implies a certain degree of disorder in Clar ring arrangements, which gives a plausible explanation to why incomplete-Clar family is insensitive to changes in topology also with respect to energy gaps.  

A complementary point of view for the changes in the energy gaps is to use the symmetry of the bands near the Fermi-level. Consider an eigenstate $\psi$ with the translational symmetry  $\D(\T)\psi=e^{-i\kappa_T}\psi$ and the roto-translational symmetry $\D(\R\T)\psi=e^{-i\kappa_{RT}}\psi$. The state $\psi$ can be labeled both by the quantum numbers $\kappa_T$ and by $\kappa_{RT}$. Because $\D(\R\T)=\D(\T)\D(\R)$, the labels for a symmetric state [$\D(\R)\psi=\psi$] obey $\kappa_T=\kappa_{RT}$ and the labels for an antisymmetric state [$\D(\R)\psi=-\psi$] obey $\kappa_T=\kappa_{RT}-\pi$. Therefore, all antisymmetric bands in M\"obius ribbons get shifted in $\kappa$-space by $\pi$, and because for Clar and Kekul\'e families the bands near the Fermi-level are antisymmetric, shifting causes the gap to change (\reffig{fig:band_gap}b). For the straight ribbons the gap occurs at the $\Gamma$-point, which is included in the sampling with all $L$, thus making the gap only modestly dependent on length (\reffig{fig:band_gap}b).

%Total energy and bang gap differences are consequences of the band shifting when the structures are identical.  
                             
%For Clar and Kekul\'e families gap difference between the two topologies is roughly inversely proportional to the aspect ratio (\reffig{fig:band_gap}b). In fact, because 

\begin{figure}[t!]
\begin{center}
	\includegraphics[width= 0.8\linewidth]{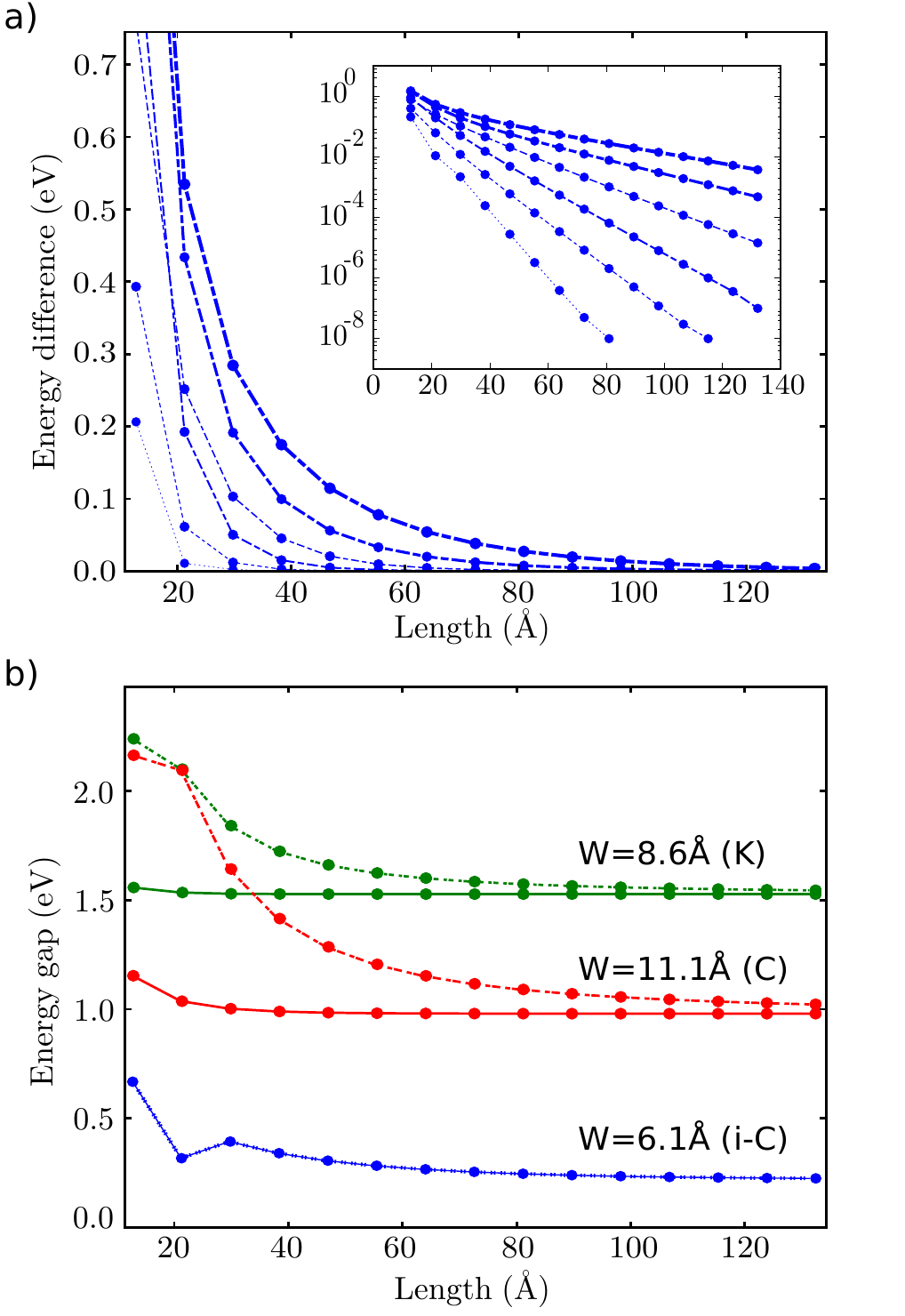}
\end{center}
	\caption{\label{fig:length_dependence} The decay of M\"obius character upon increasing aspect ratio. a) Energy difference $E_{\text{straight}} - E_{\text{M\"obius}}$ as a function of ribbon length for different widths. The linewidth is proportional to width [widths are 6.1 \r{A}  (i-C), 8.6 \r{A}  (K), 11.1 \r{A}  (C), 20.9 \r{A}  (i-C), 23.4 \r{A}  (K) and 25.8 \r{A}  (C)]. Inset: the same plot on logarithmic scale. b) Band gaps for the three narrowest i-C, K, and C structures as a function of ribbon length. Solid lines are for straight ribbons, dashed lines for M\"obius ribbons. Symbols indicate structural families.}
\end{figure}

%ENERGY DEVIATION WRTL
Regarding length-dependence, it is natural to ask that when does M\"obiusness matter? How long ribbons or how large their aspect ratios need to be for the ribbons to still be affected by changes in topology? In terms of the total energy difference between straight and M\"obius ribbons, the effect of topology seems to fall exponentially when the length increases (\reffig{fig:length_dependence}a). Meanwhile, for a given length the effect from topology is larger for wider ribbons, as expected. In overall, therefore, the effect decays exponentially with increasing aspect ratio, even though the effect appears relatively small already for the smallest realistic aspect ratios.\cite{starostin} 

The monotonous decay of the effect from M\"obius topology can be seen also in the energy gap (\reffig{fig:length_dependence}b). The decay is understood easily by inspecting the energy gap in \reffig{fig:band_gap}b as the ribbon lengthens. When the length increases---meaning larger $M$ and denser $\kappa$-point sampling---the gap will become independent of the band shifting that is caused by the change in topology. If the bands are symmetric, as they are for the icomplete-Clar family, gaps will remain independent of topology altogether.

Finally, these results prove---perhaps for the first time explicitly---that the boundary conditions do not matter after the total energy has been converged with respect to the number of $\mathbf{k}$-points in the calculation. This is a highly non-trivial observation, as a convergence with respect to the number of $\mathbf{k}$-points does not automatically mean that topology would become unimportant---topology is, after all, a conserved property whose effect cannot be monitored continuously. Although these results are for one-dimensional ribbons, they give reassurance that the bizarre, unphysical three-dimensional translational-periodic topology would become equally unimportant upon $\mathbf{k}$-point convergence. 

%We also see that the M\"obius topology is lower in energy. Hence, the M\"obius graphene nanoribbons differ from hydrocarbon annulenes, where it has been predicted that systems with $4n + 2$ pi-electrons would be antiaromatic for M\"obius topology \cite{heilbronner}.    

\section{Conclusion}

The treatment of these topological M\"obius ribbons bears certain natural concerns. The first concern, as mentioned earlier, is the absence of strain in the bond lengths. The second concern is the absense of the finiteness of the twisted structure, essential for some of the calculated effects of M\"obius ribbons, such as the spectral splitting under weak electric fields.\cite{zhao} The third concern is the constrained symmetry, the inability for symmetry breaking and for localized electronic states. The fourth concern is the absence of spin. \emph{Ab initio} calculations have shown that M\"obius zigzag graphene ribbons are stable ferromagnets.\cite{wang} This spontaneous spin-polarization of zigzag graphene nanoribbons was our excuse to exclude them.\cite{Son2006} Armchair ribbons, again, do not have such a spontaneous polarization, and provide more reasonable systems for spin-unpolarized calculations. All the same, the investigation of magnetic properties will require calculations with finite ribbons; unit cell calculations with M\"obius boundary conditions could never account for such effects.

However, the above concerns are not detrimental for our goal. The purpose of this work was to investigate the effects of topology \emph{without these other effects}, to investigate the effect of topology alone. We found that the same structural families survive the topological change, and that within each family the effects can be understood in terms of Clar ring arrangements and their rigidity. We found that purely topological effects are unexpectedly short-ranged, which means that the more important effects of M\"obius topology will arise from geometrical distortions, finite-size effects, and electron-electron interactions. Our findings also help to understand the ubiquitous phrase regarding ``$\mathbf{k}$-point convergence'' which can now be undestood to mean convergence not only with respect to the system size, but also with respect to the global, overall topology: a small unit cell with many enough repetitions \emph{soon loses its perception of the global topology}.

%While the changes in topology for real ribbons makes big differences for total energies, geometries, and electronic structures, we find that the M\"obius topology itself has an unexpectedly small effect on ribbons' properties, especially for realistic aspect ratios. <-- Siirsin taman intron tuloksiin 

\section{Acknowledgements}
We acknowledge the Academy of finland for funding. 

%\bibliographystyle{elsarticle-num}
%\bibliography{bibliography,library}

\end{document}